\documentclass[a4paper,11pt,review]{elsarticle}
\pdfoutput=1

\journal{Journal of Computational and Applied Mathematics}
\usepackage{amsmath,amssymb,natbib,graphicx,hyperref,setspace,xfrac}
\usepackage[left=1in,right=1in,top=1.4in,bottom=1.2in]{geometry}

\hypersetup{
	pdftitle={Inverse Gaussian quadrature and finite normal-mixture approximation of the generalized hyperbolic distribution},
	pdfauthor={Jaehyuk Choi, Yeda Du, Qingshuo Song},
	pdfkeywords={Normal variance-mean mixture, generalized hyperbolic distribution, inverse Gaussian distribution, numerical quadrature},
	colorlinks=true,
	linkcolor=red,
	citecolor=blue,
	urlcolor=blue,
	bookmarksnumbered=true,
	pdfstartview=
}

\newtheorem{definition}{Definition}
\newtheorem{lemma}{Lemma}
\newtheorem{theorem}{Theorem}
\newtheorem{corollary}{Corollary}
\newproof{pf}{Proof}

\newcommand{\tran}{\phi_\sigma}

\newcommand{\IG}{\textsc{ig}}
\newcommand{\GIG}{\textsc{gig}}
\newcommand{\GH}{\textsc{gh}}
\newcommand{\He}{H\!e}
\newcommand{\half}{\sfrac{1}{2}}
\newcommand{\qtext}[2][\quad]{#1\text{#2}#1}

\begin{document}
\begin{frontmatter}
\title{Inverse Gaussian quadrature and finite normal-mixture approximation of the generalized hyperbolic distribution}

\author[phbs]{Jaehyuk Choi\corref{corrauthor}}
\ead{jaehyuk@phbs.pku.edu.cn}
\author[phbs]{Yeda Du}
\author[wpi]{Qingshuo Song}
\cortext[corrauthor]{Corresponding author \textit{Tel:} +86-755-2603-0568, \textit{Address:} Peking University HSBC Business School, University Town, Nanshan District, Shenzhen 518055, China}

\address[phbs]{Peking University HSBC Business School, Shenzhen, China}
\address[wpi]{Department of Mathematical Sciences, Worcester Polytechnic Institute, Worcester, MA, USA}

\date{2 December, 2020}

\begin{abstract}
In this study, a numerical quadrature for the generalized inverse Gaussian distribution is derived from the Gauss--Hermite quadrature by exploiting its relationship with the normal distribution. The proposed quadrature is not Gaussian, but it exactly integrates the polynomials of both positive and negative orders. Using the quadrature, the generalized hyperbolic distribution is efficiently approximated as a finite normal variance--mean mixture. Therefore, the expectations under the distribution, such as cumulative distribution function and European option price, are accurately computed as weighted sums of those under normal distributions. The generalized hyperbolic random variates are also sampled in a straightforward manner. The accuracy of the methods is illustrated with numerical examples. 
\end{abstract}

\end{frontmatter}

\section{Introduction} \noindent
The inverse Gaussian (IG) distribution, $\IG(\gamma,\delta)$, has the density function
$$
f_\IG(x\,|\,\gamma,\delta) = \frac{\delta}{\sqrt{2\pi x^3}}\; \exp\left(- \frac{(\gamma x-\delta)^2}{2x}\right)\qtext{for} \gamma\ge 0,\; \delta>0.
$$
The first passage time of a drifted Brownian motion, $\gamma t + B_t$, to a level, $\delta$, is distributed by $\IG(\gamma,\delta)$. The term \textit{inverse} refers to the time of Brownian motion at a fixed location, whereas the Gaussian distribution refers to the location at a fixed time. See \citet{folks1978ig} for a review on the properties of the IG distribution. It is further extended to the generalized inverse Gaussian (GIG) distribution, $\GIG(\gamma,\delta,p)$, with density
$$ f_\GIG(x\,|\,\gamma,\delta,p) = \frac{(\gamma/\delta)^p\, x^{p-1}}{2K_p(\gamma\delta)} \exp\left( - \frac{\gamma^2 x^2+\delta^2}{2x}\right),
$$
where $K_p(\cdot)$ is the modified Bessel function of the second kind with index $p$. With $K_{-\half}(z) = \sqrt{\pi/2z}\; e^{-z}$, it can be shown that $\IG(\gamma,\delta) \sim \GIG(\gamma,\delta,-\half)$. 
The GIG random variate, $X \sim \GIG(\gamma,\delta,p)$, has the scaling property: $X \sim (\delta/\gamma)\, \GIG(\sigma,\sigma,p)$ with $\sigma=\sqrt{\gamma\delta}$. Therefore, any statement for $\GIG(\sigma,\sigma,p)$ can be easily generalized to $\GIG(\gamma,\delta,p)$. The reciprocal also follows a GIG distribution: $1/X\sim \GIG(\delta,\gamma,-p)$. See \citet{koudou2014charac} for the properties of the GIG distribution. The mean, variance, skewness, and ex-kurtosis of $\IG(\sigma,\sigma)$ are 1, $1/\sigma^2$, $3/\sigma$, and $15/\sigma^2$, respectively. Therefore, the IG (and GIG) distribution is more skewed and heavy-tailed as $\sigma$ becomes smaller.

When $X\sim\GIG(\gamma,\delta,p)$ is used as the mixing distribution of the normal variance--mean mixture,
\begin{equation} \label{eq:nvmm}
Y = \mu + \beta X + \sqrt{X} Z \quad\text{for standard normal variate}\; Z,
\end{equation}
the generalized hyperbolic (GH) variate, $Y \sim\GH(\mu,\beta,\gamma,\delta,p)$, is obtained with density
$$ f_{\GH}(y\,|\,\mu, \beta, \gamma, \delta, p) = 
\frac{\sqrt{\alpha}\,(\gamma/\alpha\delta)^p}{\sqrt{2\pi}\, K_p(\delta\gamma)} e^{\beta(y-\mu)} \frac{K_{p-\half}(\alpha \sqrt{\delta^2+(y-\mu)^2})}{(\delta^2+(y-\mu)^2)^{(1-2p)/4}},
$$
where $\alpha = \sqrt{\beta^2+\gamma^2}$.\footnote{In the literature, the GH distribution is equivalently parameterized by $\mu$, $\alpha$, $\beta$, $\delta$, and $p$ with the restriction $|\beta|<\alpha$.} The scaling property of the GIG distribution implies that the parameters of $Y \sim\GH(\mu,\beta,\gamma,\delta,p)$ can be normalized to $ \sqrt{\gamma/\delta\,}(Y - \mu) \sim \GH(0, \tilde{\beta}, \sigma, \sigma, p)$ where $\sigma = \sqrt{\gamma\delta}$ and $\tilde{\beta} = \beta\sqrt{\delta/\gamma}$. Therefore, any statement for $\GH(0, \tilde{\beta}, \sigma, \sigma, p)$ can be easily generalized to $\GH(\mu,\beta,\gamma,\delta,p).$

As the name suggests, the GH distribution generalizes the hyperbolic distribution, the $p=1$ case, originally studied for the sand particle size distributions~\citep{barndorff1977exp}. Later, the GH distribution was applied to finance~\citep{eberlein1995hyp,prause1999gh}. Particularly, the normal inverse Gaussian (NIG) distribution, the $p=-\half$ case, draws attention as the most useful case of the distribution owing to its better probabilistic properties~\citep{barndorff1997nig,barndorff1997nig_sv} and superior fit to empirical financial data~\citep{kalemanova2007normal,corlu2015fx}. The model-based clustering with the GH mixtures has recently been proposed as a better alternative to Gaussian mixtures to handle skewed and heavy-tailed data~\citep{browne2015mix}.

Despite the wide applications, the evaluation involving the GH distribution is not trivial. For example, the cumulative distribution function (CDF) has no closed-form expression, and thus must resort to the numerical integration of the density function~\citep{scott2018gh_R}, which is computationally costly. Regarding financial applications, efficient numerical procedures for pricing European option are still at large. While a closed-form solution is known for a subset of the NIG distribution~\citep{ivanov2013closed}, option pricing currently depends on the Quasi-Monte Carlo method~\citep{imai2009accel}.

This study proposes a novel and efficient method to approximate the GH distribution as a finite normal variance--mean mixture. Therefore, an expectation under the GH distribution is reduced to that under normal distribution for which analytic or numerical procedures are broadly available. The CDF and vanilla option price under the GH distribution are computed as a weighted sum of the normal CDFs and the Black-Scholes prices, respectively. The components and weights of the finite mixture are obtained by constructing a new numerical quadrature for the GIG distribution---the mixing distribution---by exploiting its relationship with the normal distribution. While the Gauss--Hermite quadrature for the normal distribution exactly evaluates positive moments only, the proposed quadrature exactly evaluates both positive and negative moments. Additionally, the new quadrature can be used as an alternative method for sampling random variates from the GH distribution (and the GIG distribution to some extent). Except for the NIG distribution~\citep{michael1976gen}, the sampling of the GH distribution depends on the acceptance--rejection methods for the GIG distribution~\citep{dagpunar1989easily,hormann2014gig_rv}. Compared to existing methods, our method based on the quadrature is more straightforward to implement and there are no rejected random numbers.

This paper is organized as follows. Section~\ref{sec:quad} discusses the numerical quadrature and its benefits for mixture distributions. Section~\ref{sec:igquad} derives the quadratures for the IG and GIG distributions. Section~\ref{sec:num} deals with numerical examples and Section~\ref{sec:con} concludes the study.

\section{Numerical quadrature for mixing distribution} \label{sec:quad} \noindent
The Gaussian quadrature with respect to the weight function $w(x)$ on the interval $(a,b)$ is the abscissas, $\{x_k\}$, and weights, $\{w_k\}$, for $k=1,\ldots,n$, that best approximate the integral of a given function $g(x)$ as
$$ \int_a^b g(x) w(x) dx \approx \sum_{k=1}^n g(x_k)\,w_k.$$
The points and weights are the most optimal in that they exactly evaluate the integral when $g(x)$ is a polynomial up to degree $2n-1$. When $w(x)$ is a probability density, the weights have the desired property: $\sum_{k=1}^n w_k=1$ from $g(x)=1$. 
It is known that $\{x_k\}$ are the roots of the $n$th-order orthogonal polynomial, $p_n(x)$, with respect to $w(x)$ and $(a,b)$, and $\{w_k\}$ are given as the integral of the Lagrange interpolation polynomial
$$ w_k = \frac{1}{p'_n(x_k)} \int_a^b \frac{p_n(x)}{x - x_k} w(x) dx.
$$
The Gaussian quadratures have been found for several well-known probability densities $w(x)$: Gauss--Legendre quadrature for uniform distribution, Gauss--Jacobi for beta distribution, and Gauss--Laguerre for exponential distribution. In particular, this study heavily depends on the Gauss--Hermite quadrature for the normal distribution. In the rest of the paper, the Gauss--Hermite quadrature is always defined with respect to the standard normal density, $w(x) = e^{-x^2/2}/\sqrt{2\pi}$, rather than $w(x) = e^{-x^2}$. Therefore, the orthogonal polynomials are the probabilists' Hermite polynomials denoted by $\He_n(x)$ in literature, not the physicists' Hermite polynomials denoted by $H_n(x)$. 

If an accurate quadrature, $\{x_k\}$ and $\{w_k\}$, were known for the mixing distribution $X$ in Eq.~\eqref{eq:nvmm}, an expectation involving $Y$ can be approximated as a finite mixture of normal distributions with mean $\mu+\beta\,x_k$ and variance $x_k$:
\begin{equation} \label{eq:exp}
\mathbb{E}\big(g(Y)\big) \approx \sum_{k=1}^{n} w_k\,\mathbb{E}\big(g(\mu + \beta x_k + \sqrt{x_k} Z)\big),
\end{equation}
for a function $g(\cdot)$ and standard normal variate $Z$. The approximated expectation can be efficiently computed because
analytic or numerical procedures are broadly available for normal distribution. For example, the CDF of the GH variate, $Y$, can be approximated as the weighted sum of those of the normal distribution
\begin{equation} \label{eq:cdf}
F_\GH(y) = \mathbb{P}(Y<y) \approx \sum_{k=1}^n w_k\, N\left( \frac{y-\mu}{\sqrt{x_k}}-\beta\sqrt{x_k} \right),
\end{equation}
where $N(\cdot)$ is the standard normal CDF. This approximation is particularly well suited for a CDF because the value monotonically increases from 0 to 1 since $w_k> 0$ and $\sum w_k=1$. If a stock price follows the log-GH distribution, the price of the European call option struck at $K$ can be approximated as a weighted sum of the Black--Scholes formulas with varying spot prices and volatilities
\begin{equation} \label{eq:bs}
\begin{gathered}
C_\GH(K) = \mathbb{E}(\max(e^Y-K,0)) \approx \sum_{k=1}^n w_k \big(F_k N(d_{k}+\sqrt{x_k}) - K N(d_k)\big), \\
\text{where}\quad F_k=e^{\mu + (\beta+\half)x_k}, \quad d_k = \frac{\log(F_k/K)}{\sqrt{x_k}} - \frac{\sqrt{x_k}}{2}.
\end{gathered}
\end{equation}
Even if the quantity of interest has no analytic expression under normal distribution, a compound quadrature can be constructed for $Y$, whose points and weights, respectively, are 
$$\{\mu + \beta x_k + \sqrt{x_k}\, z_l\} \text{ and } \{w_k\,h_l\} \qtext{for} k=1,\ldots,n, \text{ and } l=1,\ldots,m, $$
where $\{z_l\}$ and $\{h_l\}$ are the points and the weights, respectively, of the Gauss--Hermite quadrature.

The quadrature for the mixing distribution also serves as a quick and simple way to generate random variate of $Y$. The sampling of $Y$ is approximated as
\begin{equation} \label{eq:rv}
Y \approx \mu + \beta x_K + \sqrt{x_K}\, Z,
\end{equation}
where $K$ is the random index determined from a uniform random variate $U$ independent from $Z$,
$$ K = \inf \{k: \textstyle U \le w_1+\cdots+w_k,\;1\le k\le n \}. $$
Here, the construction of $K$ is to ensure that $x_K$ is a randomly selected point among $\{x_k\}$ according to the probability $\{w_k\}$: $\mathbb{P}(K=k) = \mathbb{P}(x_K = x_k) = w_k$.
Therefore, the expectation of $g(Y)$ evaluated with the simulated values of $Y$ is the same as that with the quadrature in Eq.~\eqref{eq:exp}:
\begin{align*}
\mathbb{E}\big(g(Y)\big) &\approx \mathbb{E}\left(g(\mu + \beta x_K + \sqrt{x_K}\, Z)\right)
= \mathbb{E}\left(\mathbb{E}\left(g(\mu + \beta x_K + \sqrt{x_K}\, Z)\,\Big|\,K\right)\right) \\
&= \sum_{k=1}^n w_k\,\mathbb{E}\big(g(\mu + \beta x_k + \sqrt{x_k}\, Z)\big).
\end{align*}
Note that $x_K$ can serve as a random variate for $X$, but the usage might be limited due to discreteness. The random number $Y$ sampled in Eq.~\eqref{eq:rv}, however, is continuous because $x_K$ is \textit{mixed} with $Z$. It is also possible to make antithetic variables by replacing $U$ with $1-U$. We will test the validity of the random number generation method with numerical experiments in Section~\ref{sec:num}. 

\section{IG and GIG Quadratures} \label{sec:igquad} \noindent
With the change of variable, $(\gamma x-\delta)^2/x = z^2$, the exponent of $f_\IG(x\,|\,\gamma,\delta)$ becomes that of the standard normal density in $z$. This mapping plays an important role in understanding this study as well as the previously known properties of the IG distribution. We define the mapping appropriately and derive a key lemma.
\begin{definition} \label{def} Let $\tran$ be a monotonically increasing one-to-one mapping from $x\in(0,\infty)$ to $z\in(-\infty,\infty)$, and $\tran^{-1}$ be the inverse mapping, respectively, defined as
$$ z = \tran(x) = \sigma\left(\sqrt{x} - \frac{1}{\sqrt{x}}\right) \qtext{and}
x = \tran^{-1}(z) = 1 + \frac{z^2}{2\sigma^2} + \frac{z}{\sigma} \sqrt{1 + \frac{z^2}{4\sigma^2}}.
$$
\end{definition}
\begin{lemma} \label{lemma1}
	The mapping, $z = \tran(x)$, relates the IG density, $f_\IG(x\,|\,\sigma,\sigma)$, and the standard normal density, $n(z)$, as follows:
	\begin{equation} 
	f_\IG(x\,|\,\sigma,\sigma)\, \frac{1 + x}{2}\, dx = n(z)\, dz.
	\end{equation} 
\end{lemma}
\begin{pf}
	The proof is trivial from the differentiation, 
	$$ \frac{dz}{dx} = \tran'(x) = \sigma\, \frac{1 + x}{2\sqrt{x^3}}.$$ 
	\hfill $\square$ 
\end{pf}
With Lemma~\ref{lemma1}, two important results about the IG distribution can be obtained. Let $x_+$ and $x_-$ be $x_\pm = \tran^{-1}(\pm z)$ for $z\ge 0$. Then, $x_+x_- = 1$ and $0< x_- \le 1 \le x_+$. For standard normal $Z$ and $X\sim \IG(\sigma,\sigma)$, the probability densities around the three variables, $x_+$, $x_-$, and $z$, satisfy
\begin{equation} \label{eq:pm_sum}
\mathbb{P}(X \in dx_+) + \mathbb{P}(X \in dx_-) = 
\frac{2\,\mathbb{P}(Z\in dz)}{1+x_+} + \frac{2\,\mathbb{P}(Z\in d(-z))}{1+x_-} = 2\,\mathbb{P}(Z\in dz),
\end{equation}
where
\begin{equation*}
\mathbb{P}(X \in dx_\pm) = f_\IG(x_\pm\,|\,\sigma,\sigma) dx_\pm \qtext{and}
\mathbb{P}(Z\in dz) = n(z) dz.
\end{equation*}
It follows that
\begin{align*}
\mathbb{P}\left(\tran(X)^2 < z^2\right) &= \mathbb{P}\left(x_-< X < x_+\right) 
= \int_{x_-}^1 \mathbb{P}(X \in dx_-) + \int_1^{x_+} \mathbb{P}(X \in dx_+) \\
&= \int_0^z 2\mathbb{P}(Z\in dz) = \mathbb{P}(Z^2<z^2).
\end{align*}
Thus, $\tran(X)^2 = \sigma^2 (X - 1)^2/X$ is distributed as the chi-squared distribution with 1 degree of freedom~\citep{shuster1968inv}. Eq.~\eqref{eq:pm_sum} also implies that choosing between the two random values, $X_\pm=\tran^{-1}(\pm|Z|)$, with probabilities, $p_\pm = 1/(1+ X_\pm)$~~($p_+ + p_-=1$), respectively, is an exact sampling method of $\IG(\sigma,\sigma)$~\citep{michael1976gen}, which originally provided key insight for this study. 

\begin{lemma} \label{lemma2}
Let $\{z_k\}$ and $\{h_k\}$ be the points and the weights, respectively, of the Gauss--Hermite quadrature from the $n$th-order Hermite polynomial $\He_n(z)$. Then, the points $\{x_k\}$ transformed by $x_k = \tran^{-1}(z_k)$ and the weights $\{h_k\}$ serve as a numerical quadrature with respect to $w(x)=f_\IG(x\,|\,\sigma,\sigma)\,(1+x)/2$ over the domain $(0,\infty)$. The corresponding orthogonal functions are $G_n(x) = \He_n\circ\tran(x)$. 
\end{lemma}
\begin{pf}
The following proof is a straightforward result from Lemma~\ref{lemma1}, which states that, for a function $g(x)$,
\begin{equation} \nonumber
\int_0^\infty g(x) f_\IG(x\,|\,\sigma,\sigma) \frac{1+x}{2}\, dx = \int_{-\infty}^\infty 
g\circ\tran^{-1}(z)\, n(z)\, dz.
\end{equation}
First, the functions $G_n(x)$ are orthogonal because 
$$ \int_0^\infty G_n(x) G_{n'}(x) f_\IG(x\,|\,\sigma,\sigma) \frac{1+x}{2}\, dx = \int_{-\infty}^\infty \He_n(z) \He_{n'}(z)\, n(z)\, dz = n!\, \delta_{nn'},$$
where $\delta_{nn'}$ is the Kronecker delta.
Second, $\{x_k\}$ are the roots of $G_n(x)=0$ since $G_n(x_k)=\He_n(z_k)=0$. Finally, the weight $h_k$ is invariant under the mapping $z=\tran(x)$:
$$ \frac{\tran'(x_k)}{G_n'(x_k)}\int_0^\infty \frac{G_n(x)}{\tran(x) - z_k} f_\IG(x\,|\,\sigma,\sigma) \left.\frac{1+x}{2}\right. dx = \frac{1}{\He_n'(z_k)}\int_{-\infty}^\infty \frac{\He_n(z)}{z-z_k}n(z) dz = h_k.
$$ \hfill $\square$
\end{pf}
From Lemma~\ref{lemma2}, the expectation of $g(X)$ under the IG distribution is evaluated with $\{x_k\}$ and $\{h_k\}$ as follows:
\begin{equation} \label{eq:int_trans}
\int_0^\infty g(x) f_\IG(x\,|\,\sigma,\sigma)\, dx = \int_0^\infty \frac{2g(x)}{1+x} f_\IG(x\,|\,\sigma,\sigma) \frac{1+x}{2}\, dx = \sum_{k=1}^{n} g(x_k) \frac{2 h_k}{1+x_n}
\end{equation}
This observation leads us to the numerical quadrature with respect to the IG distribution.
\begin{theorem}[\textbf{IG Quadrature}] \label{theorem}
	Let $\{z_k\}$ and $\{h_k\}$ be the points and the weights, respectively, of the Gauss--Hermite quadrature from the $n$th-order Hermite polynomial $\He_n(z)$. Then, the points $\{x_k\}$ and the weights $\{w_k\}$, defined by
	$$ x_k = \frac{\delta}{\gamma}\;\tran^{-1}(z_k) \qtext{and} w_k = \frac{2\,h_k}{1+\tran^{-1}(z_k)} \qtext{for} \sigma = \sqrt{\gamma\delta}, $$
	serve as a numerical quadrature with respect to $w(x)=f_\IG(x\,|\,\gamma,\delta)$ over the domain $(0,\infty)$. The quadrature exactly evaluates the $r$th-order moments for $r=1-n,\ldots,n$.
\end{theorem}
\begin{pf}
	Thanks to the scaling property of the GIG random variate, it is sufficient to consider the case $\gamma=\delta=\sigma$.
	The construction of the new weights $\{w_k\}$ immediately follows from Eq.~\eqref{eq:int_trans}. We need to prove the statement about the moments:
	$$ \mathbb{E}(X^r) = \sum_{k=1}^n x_k^r\, w_k. $$
	The change in variable, $y=1/x$ yields $ f_\IG (x\,|\,\sigma,\sigma) dx = -y\, f_\IG(y\,|\,\sigma,\sigma) dy$ and $\mathbb{E}(X^r) = \mathbb{E}(X^{1-r})$ for $X \sim \IG(\sigma,\sigma)$.\footnote{See Eq.~\eqref{eq:mom} for the analytic expression of the moments ($p=-\half$). The property, $\mathbb{E}(X^r) = \mathbb{E}(X^{1-r})$, can be directly proved with the symmetry, $K_p(\cdot) = K_{-p}(\cdot)$.} Therefore, the left-hand side is expressed as
	$$ \mathbb{E}(X^r) = \frac12 \mathbb{E}(X^r+X^{1-r}) = \mathbb{E}\left(\frac{1+X}{2} \theta_r(X)\right) = \mathbb{E}\left(\theta_r\circ \tran^{-1}(Z)\right).
	$$
	where $\theta_1(x)=1$ and 
	$$ \theta_r(x) = \frac{x^r + x^{1-r}}{1+x} = (-1)^{r-1} + \sum_{j=1}^{r-1}(-1)^{r-1-j}\left(x^j+\frac{1}{x^j}\right) \qtext{for} r\ge 2. $$
	The quadrature integration on the right-hand side also satisfies a similar property, $\sum_{k=1}^n x_k^r\, w_k = \sum_{k=1}^n x_k^{1-r}\, w_k$, because of the symmetry of the quadrature points, $1/x_k = \tran(-z_k)$. Therefore, the right-hand side is expressed as
	$$ \sum_{k=1}^n x_k^r\, w_k = \sum_{k=1}^n \frac{1+x_k}{2}\,\theta_r(x_k) w_k = \sum_{k=1}^n \theta_r\circ\tran^{-1}(z_k)\, h_k.
	$$
	For the two sides to be equal, the Gauss--Hermite quadrature integration of $\theta_r\circ\tran^{-1}(z)$ should be exact and this is the case if $\theta_r\circ\tran^{-1}(z)$ is a polynomial of $z$ of degree $2n-1$ or below. It can be shown using Chebyshev polynomials. If $T_j(\cdot)$ is the $j$th-order Chebyshev polynomials of the first kind, then it has a property, $T_j(\cosh(y))=\cosh(jy)$. With the changes of variables, $x = e^y$ and $z=\tran(x)$, we can express 
	$$x^j+\frac{1}{x^j} = 2\,\cosh(jy) = 2\,T_j(\cosh(y)) = 2\,T_j\!\left(\frac{z^2}{2\sigma^2}-1\right).$$
	Therefore, $\theta_r\circ\tran^{-1}(z)$ is a linear combination of $T_j(z^2/(2\sigma^2)-1)$ for $j=0,\ldots,r-1$, thereby an order $2(r-1)$ polynomial of $z$. It follows that the quadrature integration of the $r$th-order moment is exact for $r=1,\ldots,n$. From the symmetry $\mathbb{E}(X^r) = \mathbb{E}(X^{1-r})$, the same holds for $r=1-n,\ldots,0$. \hfill $\square$
\end{pf}
The following remarks can be made on the new quadrature. First, the orthogonal functions, $G_n(x)$, are not polynomials of $x$; therefore, the quadrature is not a Gaussian quadrature. Given below are first a few orders of $G_n(x)$ for $\IG(1,1)$ obtained from $\He_n(z)$:
$$
\begin{aligned}
G_0(x) &= 1, \quad & \He_0(z)&=1 \\
G_1(x) &= \frac{x-1}{\sqrt{x}}, \quad & \He_1(z)&=z \\
G_2(x) &= \frac{x^2-3x+1}{x},\quad & \He_2(z) &= z^2-1 \\
G_3(x) &= \frac{(x-1)(x^2-4x+1)}{x\sqrt{x}},\quad & \He_3(z) &= z^3-3z \\
\end{aligned}
$$
Nevertheless, the quadrature is accurate for integrating both positive and negative moments. Second, we name the quadrature as \textit{inverse Gaussian quadrature} after the name of the distribution. Here, the term \textit{inverse} additionally conveys the meaning that it is not a Gaussian quadrature and can accurately evaluate the \textit{inverse} moments. Third, the construction of the quadrature is intuitively understood as the method described by \citet{michael1976gen} applied to the \textit{discretized} normal random variable, $\{z_k\}$ with probabilities $\{h_k\}$, instead of the continuous normal variate. Fourth, from Lemma~\ref{lemma2}, the error estimation of the IG quadrature is obtained as a modification from that of the Gauss--Hermite quadrature~\citep[p.~890]{abramowitz}:
$$
\int_0^\infty g(x) f_\IG(x\,|\,\sigma,\sigma) dx - \sum_{k=1}w_k\,g(x_k) = \frac{n!}{(2n)!}\, H^{(2n)} (\xi) \qtext{for some} \xi \in (-\infty, \infty), 
$$
where the function $H(z)$ is 
$$ H(z) = \frac{2\, g\circ\tran^{-1}(z)}{1+\tran^{-1}(z)}.$$ Therefore, exponential convergence on $n$ is expected if $g(x)$ is an analytic function.
Lastly, the quadrature calculation is very fast since it is a mere transformation from the Gauss--Hermite quadrature, which is available from standard numerical libraries or pre-computed values.

Since the density functions, $f_{\IG}(x\,|\,\gamma,\delta)$ and $f_{\GIG}(x\,|\,\gamma,\delta,p)$, are related by
$$ f_{\GIG}(x\,|\,\gamma,\delta,p) = c(\gamma,\delta,p)\, x^{p+\half} f_{\IG}(x\,|\,\gamma,\delta) \qtext{where} c(\gamma,\delta,p) = \sqrt\frac{\pi}{2}\;\frac{\gamma^p}{\delta^{p+1}}\;\frac{e^{-\gamma\delta}}{K_p(\gamma\delta)},
$$
we can further generalize the quadrature to the GIG distribution. 
\begin{corollary}[\textbf{GIG Quadrature}] \label{cor:gig}
	Let $\{x_k\}$ and $\{w_k\}$ be the IG quadrature with respect to $f_\IG(x\,|\,\gamma,\delta)$ defined in Theorem~\ref{theorem}.
	Then, $\{x_k\}$ and $\{\bar{w}_k\}$ defined by $\bar{w}_k = c(\gamma,\delta,p)\; x_k^{p+\half}\; w_k$ serve as a quadrature with respect to $f_\GIG(x\,|\,\gamma,\delta,p)$.
	The quadrature exactly evaluates the $r$th-order moment for $r=1-n-\alpha,\ldots,n-\alpha$ for $\alpha = p + \half$.
\end{corollary}
\begin{pf}
	The modified weights $\{\bar{w}_k\}$ are obtained from $\mathbb{E}(g(\bar{X})) = \mathbb{E}\left( c(\gamma,\delta,p)X^{p+\half}g(X)\right)$ for a function $g(x)$, where $\bar{X}\sim\GIG(\sigma,\sigma,p)$ and $X\sim\IG(\gamma,\delta)$. The statement about the moments is also a direct consequence of the relation, $\mathbb{E}(\bar{X}^{r}) = c(\gamma,\delta,p)\,\mathbb{E}(X^{r+\alpha})$. \hfill $\square$
\end{pf}
Note that if $\alpha$ is not an integer, $\sum_{k=1}^{n} \bar{w}_k=1$ is not guaranteed; therefore, it is recommended to scale $\{\bar{w}_k\}$ by the factor of $1/\sum_{k=1}^n \bar{w}_k$ to ensure $\sum_{k=1}^n \bar{w}_k = 1$. However, the amount of the adjustment is very small if $|p| \ll n$ as shown in the next section.

\section{Numerical examples} \label{sec:num} \noindent
We test the IG and GIG quadratures numerically. The methods are implemented in R (Ver. 3.6.0, 64--bit) on a personal computer running the Windows 10 operating system with an Intel core i7 1.9 GHz CPU and 16 GB RAM.

\begin{figure}
	\caption{\label{fig:err_mom}
		The $\log_{10}$ of the relative error in the $r$th-order moment of $X\sim \IG(1,1)$ computed with the quadrature size $n=10$ (left) and 20 (right). The solid line (blue) denotes the positive error and the dashed line (red) denotes the negative error. The negative moments are omitted owing to the symmetry $\mathbb{E}(X^{1-r})=\mathbb{E}(X^r)$.} \vspace{2ex}
	\includegraphics[width=0.48\linewidth]{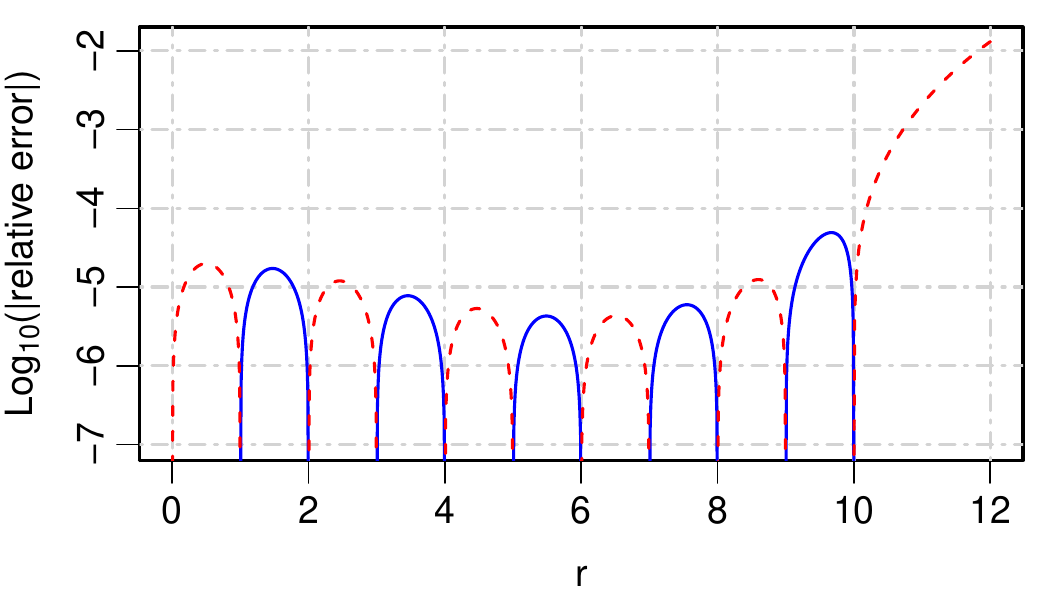} \hspace{1ex}
	\includegraphics[width=0.48\linewidth]{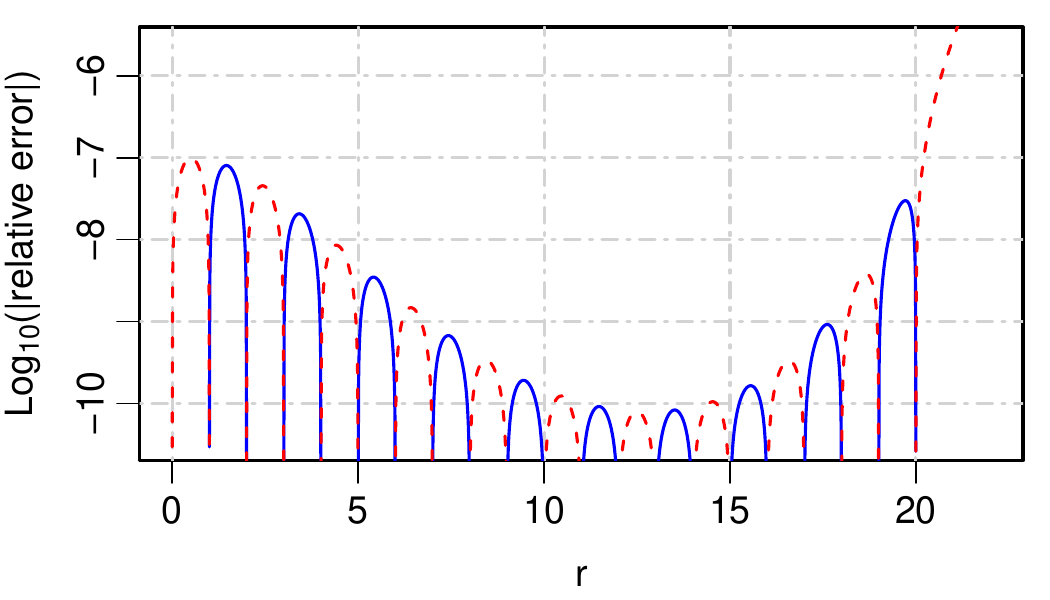}
\end{figure}

First, we evaluate the moments of the IG distribution. The $r$th-order moment of $\bar{X}\sim\GIG(\gamma,\delta,p)$ has a closed-form expression
\begin{equation} \label{eq:mom}
\mathbb{E}(\bar{X}^r) = \left(\frac{\delta}{\gamma}\right)^r\frac{K_{r+p}(\gamma\delta)}{K_p(\gamma\delta)},
\end{equation} 
against which the error of the quadrature evaluation can be measured. Figure~\ref{fig:err_mom} shows the relative error of $\mathbb{E}(X^r)$ for $X\sim\IG(1,1)$ when evaluated with $n=10$ and $20$ quadrature points. As Theorem~\ref{theorem} predicts, the quadrature exactly evaluates the moments for integer $r$ from $1-n$ to $n$. The error for non-integer $r$ is also reasonably small when $1-n\le r \le n$. The relative error of $\mathbb{E}(X^r)$ can also be interpreted as the deviation of $\mathbb{E}(\bar{X}^0=1)=\sum_{k=1}^n \bar{w}_k$ from 1 for $\bar{X}\sim\GIG(1,1,r-\half)$; thus, the sum of the GIG quadrature weights is very close to 1 if $|p|\ll n$.

\begin{figure}
	\caption{\label{fig:err_mgf}
		The convergence of the GIG distribution's MGF computed with Eq.~\eqref{eq:mgf_quad} as functions of the quadrature size $n$. The MGF is evaluated at $t=0.4\sigma^2$ (80\% of the convergence radius) for $\bar{X}\sim \GIG(\sigma,\sigma,p)$ for varying $\sigma$ values with $p=-0.5$ (upper panel), $p=1$ (middle panel), and $p=90$ (lower panel). Exact MGF is available in Eq.~\eqref{eq:mgf}.
	} \vspace{2ex}
	\centering
	\includegraphics[width=0.6\linewidth]{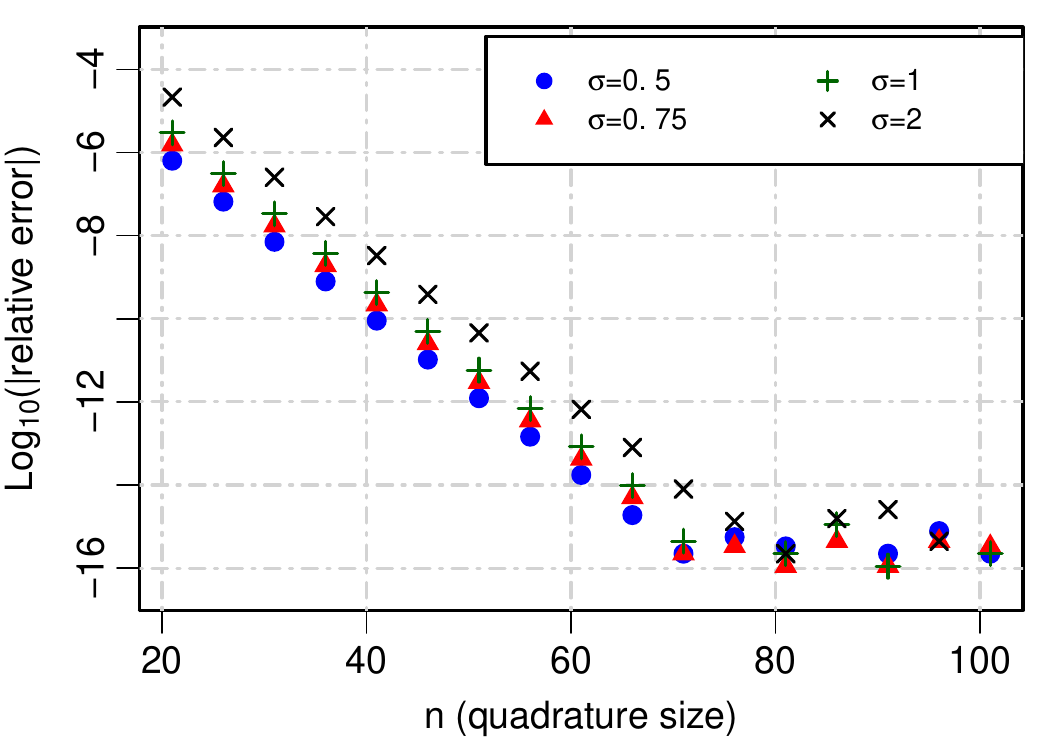} \\ \vspace{1ex}
	\includegraphics[width=0.6\linewidth]{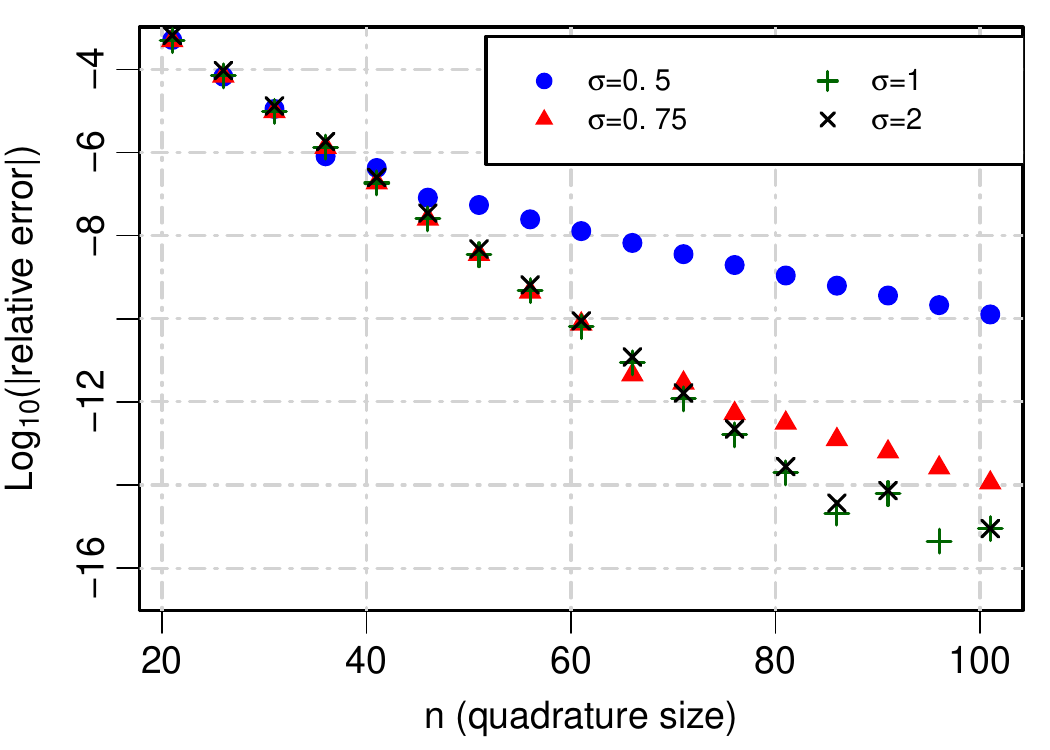} \\ \vspace{1ex}
	\includegraphics[width=0.6\linewidth]{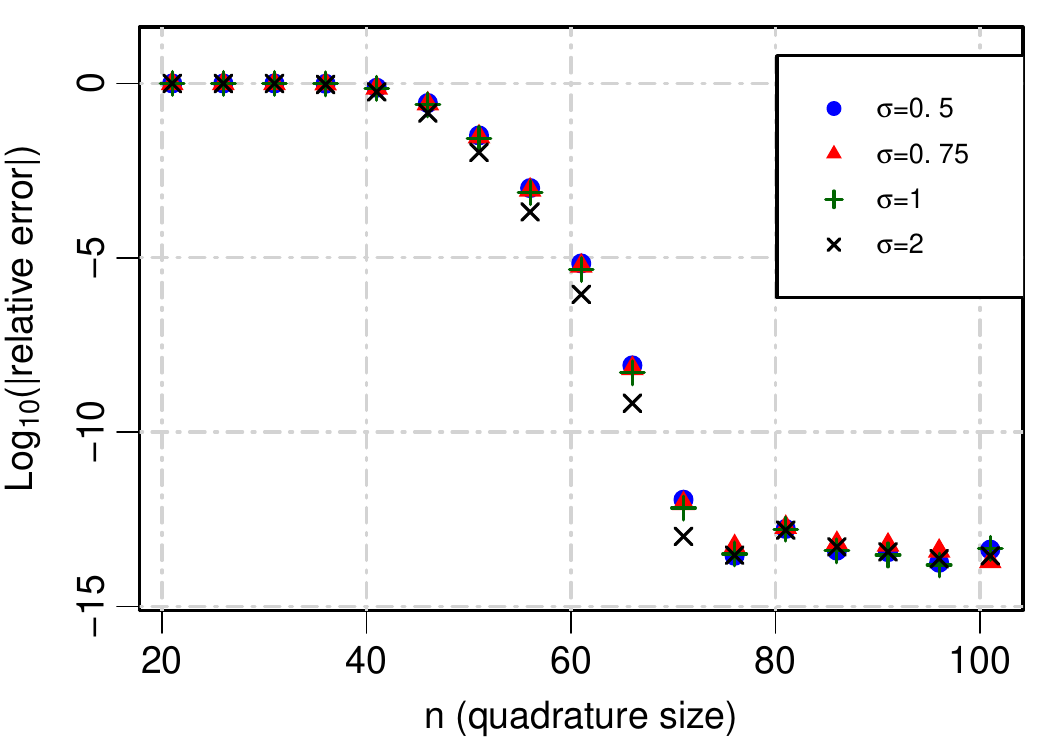} 
\end{figure}

Second, we test the accuracy of the moment generating function (MGF) of the GIG distribution. The error of the quadrature approximation is easily measured since the MGF of $\bar{X}\sim \GIG(\gamma,\delta,p)$ is analytically given by
\begin{equation} \label{eq:mgf}
M_{\bar{X}}(t) = \left(\frac{\gamma^2}{\gamma^2 - 2t}\right)^{\!p/2} \frac{K_p(\delta\sqrt{\gamma^2-2t})}{K_p(\delta\gamma)}.
\end{equation}
The MGF can be numerically evaluated with 
$M_{\bar{X}}(t) \approx \sum_{k=1}^n \bar{w}_k \exp(t\,x_k)$
for the GIG quadrature, $\{x_k\}$ and $\{\bar{w}_k\}$, from Corollary~\ref{cor:gig}. We, however, find that the numerical approximation is more accurate for negative $p$ than for positive $p$ because the probability density is more concentrated near $\bar{X}=0$ when $p<0$. Taking advantage of the symmetry, $K_p(\cdot) = K_{-p}(\cdot)$, we evaluate the MGF in a modified way for $p>0$:
\begin{equation} \label{eq:mgf_quad}
M_{\bar{X}}(t) \approx \left(\frac{\gamma^2}{\gamma^2 - 2t}\right)^{\!\max(p,0)}\sum_{k=1}^n \bar{w}_k \exp(t\,x_k),
\end{equation}
where $\{x_k\}$ and $\{\bar{w}_k\}$ are the GIG quadrature for $\bar{X}\sim \GIG(\gamma,\delta,-|p|)$.

This is a good test example to observe the convergence behavior since the MGF contains all powers of the random variable. Moreover, the MGF of the GH distributions is similarly given by function composition,
$M_{Y}(t) = \exp(\mu t)\;M_{\bar{X}}\!\left(\beta t + t^2/2\right)$. Therefore, we can also infer the accuracy of GH distribution's MGF from the result of this test. Figure~\ref{fig:err_mgf} shows the relative error of the MGF for $\bar{X}\sim \GIG(\sigma,\sigma,p)$ for $\sigma$ varying from 0.5 to 2. The error is measured at $t=0.4\,\sigma^2$, which is at the 80\% radius of the convergence radius $|t|=0.5\,\sigma^2$ when $\gamma=\delta=\sigma$. For $p$, we use the two important special cases, NIG distribution ($p=-0.5$) and hyperbolic distribution ($p=1$), and one extreme case ($p=90$).
The $p=-0.5$ case clearly shows the exponential decay of the error as functions of the quadrature size $n$, regardless of $\sigma$ values. In the $p=1$ case, however, the convergence becomes slower when $\sigma$ is smaller. This seems to be related to the fact that the orders of moments for which the GIG quadrature is exact are non-integer values ($r=\pm 0.5, \pm 1.5, \cdots$) and that the GIG distribution is more leptokurtic when $\sigma$ is smaller. In the $p=90$ case, the error quickly converges to the machine epsilon around $n\approx |p|$ after slow convergence in small $n$. The convergence pattern for $p=-90$ is very similar because of the evaluation method, Eq.~\eqref{eq:mgf_quad}. 

\begin{table}[ht]
	\caption{Parameter sets of the GH distribution for numerical experiments and their statistical properties.
	\label{tab:params}}
	\begin{center}
		\begin{tabular}{|c||c|c|c|c|} \hline
			Parameter & Set 1 & Set 2 & Set 3 & Set 4 \\ \hline\hline
$\mu$ & 0 & 0.00029 & 0.000666 & 0.000048 \\
$\alpha = \sqrt{\beta^2+\gamma^2}$ & 1 & 138.78464 & 214.4 & 9 \\
$\beta$ & 0 & $-$4.90461 & $-$6.17 & 2.73 \\
$\delta$ & 1 & 0.00646 & 0.0022 & 0.0161 \\
$p$ & $-$0.5 & $-$0.5 & 0.8357 & $-$1.663 \\ \hline \hline
$\sigma=\sqrt{\gamma\delta}$ & 1 & 0.9466 & 0.6866 & 0.3716 \\
$\tilde{\beta}=\beta\sqrt{\delta/\gamma}$ & 0 & -0.0335 & -0.0198 & 0.1183 \\
			\hline \hline
mean & 0 & 6.16E-5 & 4.00E-4 & 5.47E-4 \\
variance & 1 & 4.66E-5 & 4.33E-5 & 1.84E-4 \\
skewness & 0 & $-$0.112 & $-$0.110 & 0.655 \\
ex-kurtosis & 3 & 3.365 & 2.731 & 20.698 \\
			\hline
		\end{tabular} \\ \vspace{1em}
	\end{center}
\end{table}

Third, we evaluate the CDF of the GH distribution using Eq.~\eqref{eq:cdf}. We use the \texttt{GeneralizedHyperbolic} R package~\citep{scott2018gh_R} for a benchmark. The \texttt{pghyp} function in the package numerically integrates the probability density by internally calling the general-purpose \texttt{integrate} function\footnote{\url{https://www.rdocumentation.org/packages/stats/versions/3.6.2/topics/integrate}}, which uses adaptive quadrature. The error of the \texttt{pghyp} function is controlled by the \texttt{intTol} parameter which is, in turn, passed to the \texttt{integrate} function. We use the CDF values obtained with \texttt{intTol}=1E\,-14 as exact values.
\begin{figure}[ht]
	\caption{The convergence of the GH distribution CDF computed with the quadrature method, Eq.~\eqref{eq:cdf}, as functions of the quadrature size $n$. The error is measured as the maximum deviation on the 99 percentiles, $\{y_j=F_\GH^{-1}(j/100):\,j=1,\cdots,99\}$. \label{fig:err_cdf}}
	\centering
	\includegraphics[width=0.6\linewidth]{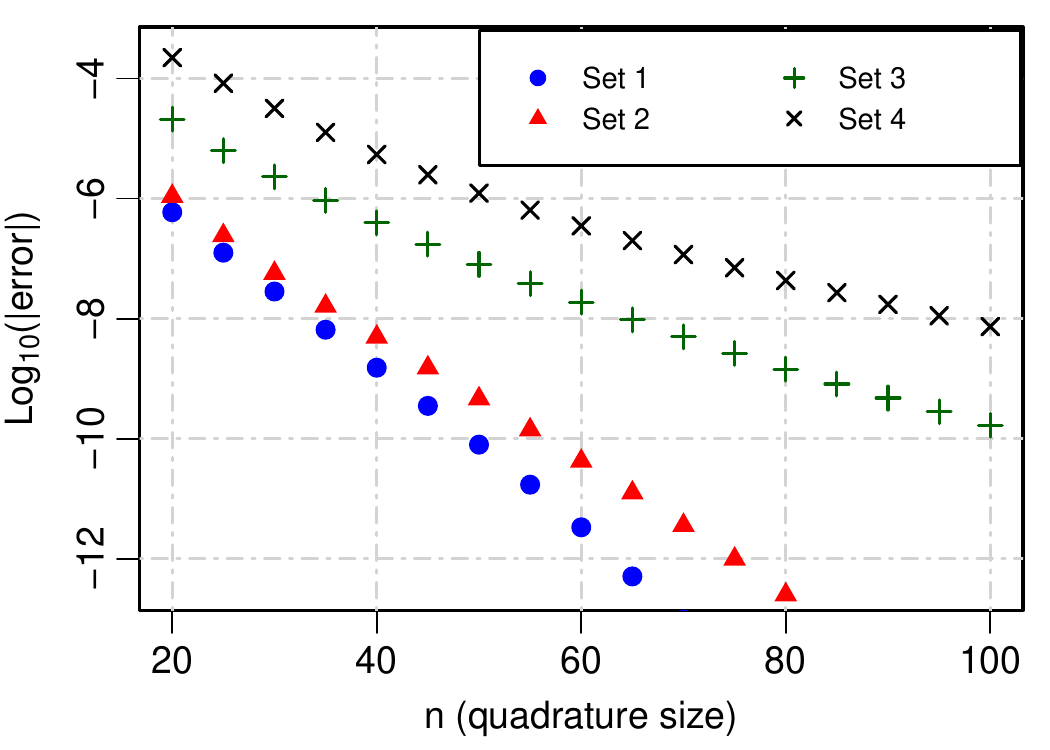} 
\end{figure}

In Table~\ref{tab:params}, we show the four parameter sets to test and their summary statistics. Set 1 is the standard NIG distribution, $\GH(0,0,1,1,-\half)$, for reference, while the rest are the parameters estimated from empirical finance data in previous studies; Set 2 is from the EUR/USD foreign exchange rate return~\citep{corlu2015fx}, and Set 3 and 4 are from the returns of the NYSE composite index and the BMW stock, respectively~\citep{prause1999gh}. 

\begin{figure}
	\caption{\label{fig:err_cdf2}
	The error of the GH distribution CDF computed with the quadrature method, Eq.~\eqref{eq:cdf}, as functions of parameters for $n=60$, 80, and 100.
	For $Y \sim\GH(0,\tilde{\beta},\sigma,\sigma,p)$, we vary $\tilde{\beta}$ (upper panel), $\sigma$ (middle panel), and $p$ (lower panel) from Set 1 ($\tilde{\beta}=0$, $\sigma=1$, and $p=-\half$). 
	The error is measured as the maximum deviation on the 99 percentiles, $\{y_j=F_\GH^{-1}(j/100):\,j=1,\cdots,99\}$. In the upper panel, the result for negative $\tilde{\beta}$ is omitted owing to the symmetry.} \vspace{2ex}
	\centering
	\includegraphics[width=0.6\linewidth]{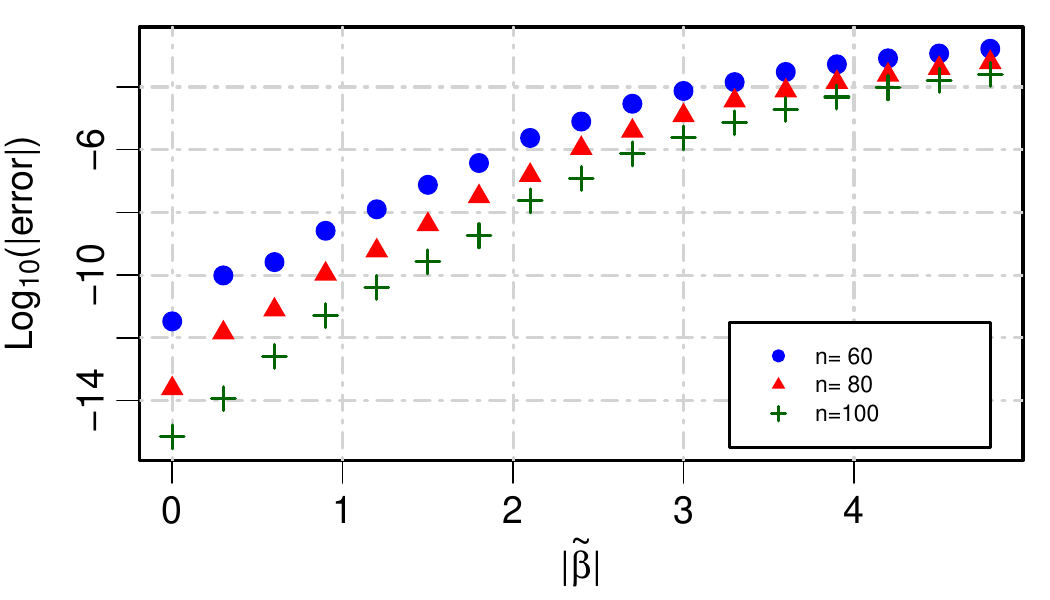} \\ \vspace{1ex}
	\includegraphics[width=0.6\linewidth]{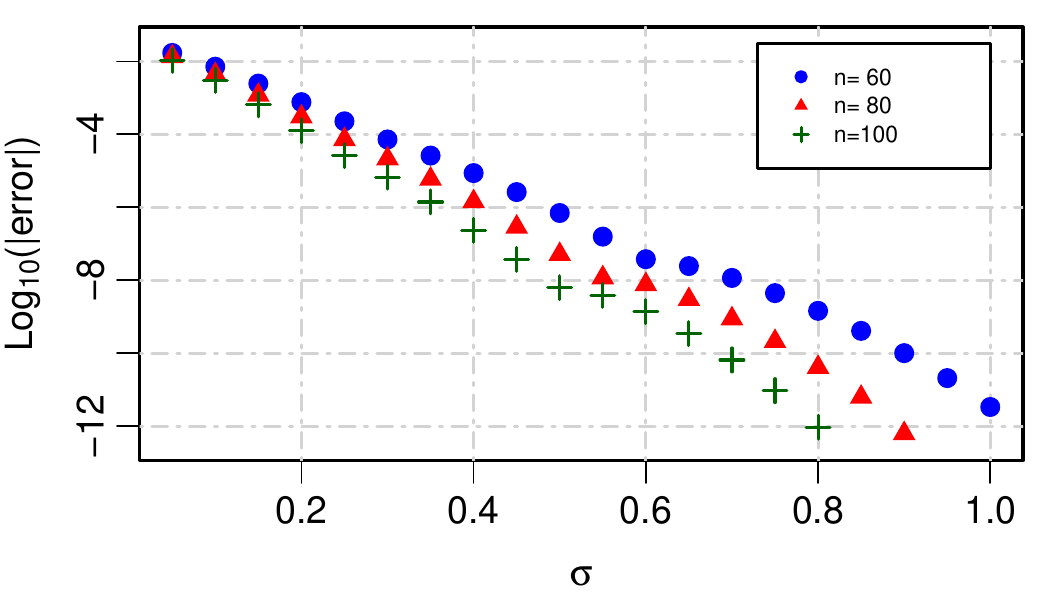} \\ \vspace{1ex}
	\includegraphics[width=0.6\linewidth]{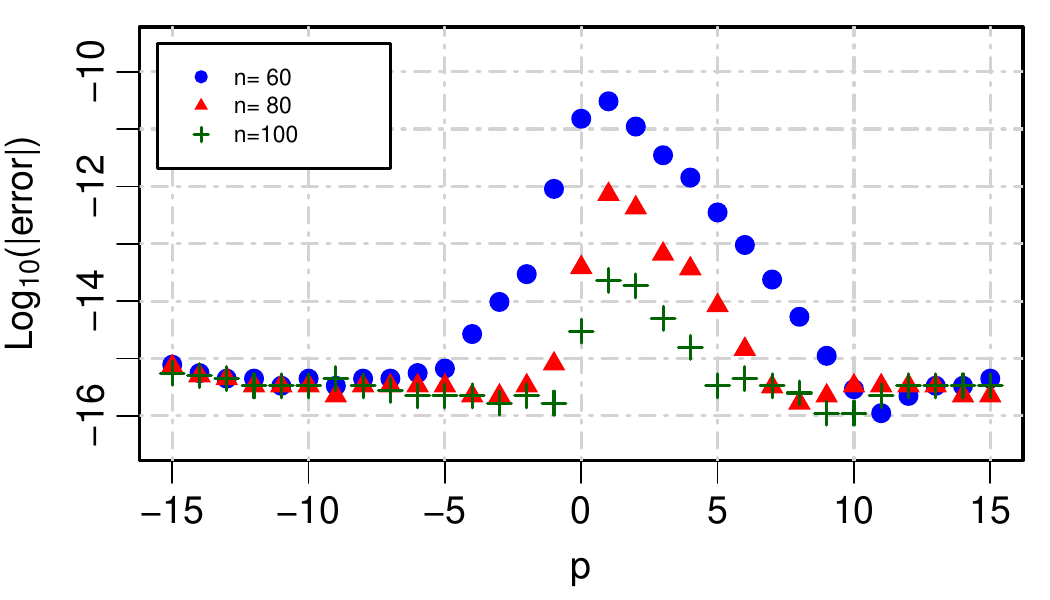} 
\end{figure}

Figure~\ref{fig:err_cdf} depicts the decay of the quadrature method error as the quadrature size $n$ increases. The error is defined as the maximum absolute deviation of the CDF values across all percentiles, $\{y_j=F_\GH^{-1}(j/100):\,j=1,\cdots,99\}$. Although the error tends to increase as $\sigma$ becomes smaller, it quickly converges to $10^{-8}$ or below around $n=100$ for all test sets. 
In Figure~\ref{fig:err_cdf2}, we additionally investigate the accuracy as functions of the distribution parameters. We similarly measure the CDF error for the normalized form, $Y \sim\GH(0,\tilde{\beta},\sigma,\sigma,p)$, when each of $\tilde{\beta}$, $\sigma$, and $p$ are varied from the values in Set 1 ($\tilde{\beta}=0$, $\sigma=1$, and $p=-\half$). Figure~\ref{fig:err_cdf2} shows that the accuracy deteriorates as $|\beta|$ becomes larger (upper panel) or $\sigma$ becomes smaller (middle panel). Therefore, the quadrature size $n$ should be larger for such parameter ranges. This also explains the convergence pattern observed in Figure~\ref{fig:err_cdf}; the convergence speed for the four sets is mainly governed by $\sigma$ since the values of $\tilde{\beta}$ are small for all cases.
Whereas, the lower panel shows that our quadrature method performs better as the GH distribution deviates away from the NIG ($p=-0.5$) and hyperbolic ($p=1$) distributions in terms of the $p$ value. This is consistent with the observation from Figure~\ref{fig:err_mgf} (lower panel).

\begin{table}[ht]
\caption{Computation time for the GH distribution CDF from the GIG quadrature, Eq.~\eqref{eq:cdf}, and the density integration~\citep{scott2018gh_R}. We measure time (in milliseconds) to compute the CDFs at the 99 percentiles. \label{tab:cpu_time}} \vspace{1ex}
\centering
\begin{tabular}{|c|c||c|c|c|c|} \hline
	Method & & Set 1 & Set 2 & Set 3 & Set 4 \\ \hline\hline
	Density integration & Error &	7.55E-08 & 3.45E-06 & 4.45E-06 & 2.56E-06 \\
	(\texttt{intTol}=2E-3) & CPU Time (ms) & 26.28 & 42.67 & 53.16 & 41.5 \\ \hline
	GIG quadrature & Error &	7.99E-11 & 4.68E-10 & 8.06E-08 & 1.24E-06 \\
	($n=50$) & CPU Time (ms) & 0.86	& 0.75 & 0.81 & 0.98 \\ \hline
\end{tabular}
\end{table}

Table~\ref{tab:cpu_time} compares the computation time of the quadrature method to that of the numerical density integration. For fair comparison, we relax the error tolerance so that the \texttt{pghyp} function runs faster. Specifically, \texttt{intTol}=2E\,-3 is chosen so that the density integration is less accurate across all parameter sets. Despite the setting, the result shows that the quadrature method is faster than the density integration at least by an order of magnitude. The performance is improved because the quadrature method avoids the expensive evaluations of the modified Bessel function, $K_p(\cdot)$.
Additionally, Table~\ref{tab:quad} reports the error in CDF at both tails. The quadrature method accurately captures the tail events. 

\begin{table}[ht]
\caption{The error of the GH distribution CDF computed with the quadrature method, Eq.~\eqref{eq:cdf}, at several extreme quantiles, $y=F_\GH^{-1}(q)$. The quadrature size, $n=50$, is used.\label{tab:quad}}
\begin{center}
	\begin{tabular}{|c||r|r|r|r|} \hline
		$q$ & Set 1\;\; & Set 2\;\; & Set 3\;\; & Set 4\;\; \\ \hline\hline
		$10^{-9}$ & 2.1E-17 & -1.4E-16 & 5.7E-17 & 4.6E-13 \\
		$10^{-6}$ & 3.8E-13 & 8.8E-14 & 1.9E-13 & 1.5E-10 \\
		$10^{-3}$ & 1.7E-10 & -1.5E-10 & 1.0E-09 & 6.3E-07 \\
		$1 - 10^{-3}$ & -1.7E-10 & -4.8E-10 & -2.9E-09 & 6.4E-06 \\
		$1 - 10^{-6}$ & -3.8E-13 & -6.8E-13 & 4.2E-13 & -1.5E-09 \\
		$1 - 10^{-9}$ & -2.1E-17 & -2.0E-16 & 1.1E-16 & 3.1E-14 \\
		\hline
	\end{tabular} \\ \vspace{1em}
\end{center}
\end{table}

\begin{table}
	\caption{The bias and standard deviation of the GH distribution CDF values computed with the Monte-Carlo method at several percentile points. The GIG random variates are generated from (a) the GIG quadrature method, Eq.~\eqref{eq:rv}, with $n=50$ and (b) the \texttt{GIGrvg::rgig} R function~\citep{leydold2017GIGrvg_R}. The CDF values are obtained from $10^6$ random numbers and the statistics are obtained after repeating 1000 simulation sets. Antithetic method is not applied. The reported values are in the unit of $10^{-6}$. \label{tab:mc}}
	\begin{center}
	(a) \begin{tabular}[t]{|c||r|r|r|r|} \hline
			Percentile & Set 1\;\;\; & Set 2\;\;\; & Set 3\;\;\; & Set 4\;\;\; \\ \hline\hline
\;\;1st & 1 $\pm$ 100 & 1 $\pm$ \ 98 & 1 $\pm$ \ 99 & 3 $\pm$ \ 99 \\
10th & 0 $\pm$ 300 & -2 $\pm$ 299 & -6 $\pm$ 298 & -3 $\pm$ 300 \\
30th & -8 $\pm$ 449 & -4 $\pm$ 447 & -3 $\pm$ 444 & -5 $\pm$ 444 \\
50th & -16 $\pm$ 513 & -14 $\pm$ 514 & -16 $\pm$ 511 & -11 $\pm$ 512 \\
70th & -19 $\pm$ 463 & -21 $\pm$ 458 & -13 $\pm$ 458 & -17 $\pm$ 456 \\
90th & -10 $\pm$ 294 & -13 $\pm$ 295 & -2 $\pm$ 298 & -4 $\pm$ 298 \\
99th & 1 $\pm$ 100 & 1 $\pm$ \ 99 & -2 $\pm$ \ 99 & 0 $\pm$ \ 98 \\
			\hline
		\end{tabular} \\ \vspace{1em}
	
	(b) \begin{tabular}[t]{|c||r|r|r|r|} \hline
			Percentile & Set 1\;\;\; & Set 2\;\;\; & Set 3\;\;\; & Set 4\;\;\; \\ \hline\hline
\;\; 1st & -1 $\pm$ 101 & -4 $\pm$ 101 & -1 $\pm$ 97 & 1 $\pm$ 103 \\
10th & 6 $\pm$ 296 & 5 $\pm$ 291 & -13 $\pm$ 306 & -2 $\pm$ 288 \\
30th & 1 $\pm$ 459 & 6 $\pm$ 454 & -24 $\pm$ 455 & -20 $\pm$ 452 \\
50th & -12 $\pm$ 505 & -19 $\pm$ 485 & -46 $\pm$ 512 & -37 $\pm$ 499 \\
70th & 13 $\pm$ 473 & 2 $\pm$ 454 & -24 $\pm$ 459 & -24 $\pm$ 459 \\
90th & -21 $\pm$ 299 & 2 $\pm$ 295 & -19 $\pm$ 301 & -15 $\pm$ 297 \\
99th & -4 $\pm$ \ 99 & 4 $\pm$ \ 99 & -6 $\pm$ 103 & -2 $\pm$ \ 99 \\
	\hline
	\end{tabular}
	\end{center}
\end{table}

Last, we test the random number generation method, Eq.~\eqref{eq:rv}. With the generated GH random variates, we evaluate the CDF values at several percentiles. In Table~\ref{tab:mc}, we report the bias\footnote{The bias is similarly measured from the \texttt{GeneralizedHyperbolic::pghyp} function with \texttt{intTol}=1E\,-14.} and standard deviation of the CDF values measured in this manner. For a benchmark, we use the \texttt{GIGrvg} R package~\citep{leydold2017GIGrvg_R} as an alternative way of generating the GIG random variate. The \texttt{rgig} in the package implements the two acceptance--rejection algorithms of \citet{dagpunar1989easily} and \citet{hormann2014gig_rv}, and optimally selects one based on the parameters. From the numerical results in Table~\ref{tab:mc}, we did not find evidence that the quadrature method is more biased than the \texttt{GIGrvg::rgig} function. While it takes 98.1 milliseconds for \texttt{GIGrvg} package to generate $10^6$ GIG random numbers on average, it takes 57.4 milliseconds for the quadrature method. 

\section{Conclusion} \label{sec:con} \noindent
The GH distribution is widely used in applications, but the expectation involving the distribution has been numerically challenging. This study shows that the GH distribution can be approximated as a finite normal mixture, and that the expectation is reduced to that of the normal distribution. For the finite mixture components, we construct novel numerical quadratures for the GIG distributions, the mixing distribution of the GH distribution. The new GIG quadrature is derived from the Gauss--Hermite quadrature. We demonstrate the accuracy and effectiveness of the method with numerical examples.

\section*{Acknowledgments}
We thank two anonymous reviewers for their helpful comments.

\singlespacing
\bibliography{../../@Bib/SV_Z2}
\end{document}